# Temperature bistability of charged dust particle in plasma

L.I.Ognev

RRC « Kurchatov institute », Institute of nuclear fusion,
Kurchatov squ.,1, Moscow, 123182, Russia

## Summary

By method of numerical simulation of heating of a dust conducting particle in homogeneous plasma it is shown, that depending on initial temperature of a particle both heating and cooling of a particle is possible with formation of two various quasi-stable conditions - *temperature bistability*. The upper condition ("hot") condition corresponds to positive charge of a particle whereas the lower ("cold") condition corresponds to negative charge of a particle. The method of experimental observation of the effect is proposed.

## Introduction

In the previous works of the author feature of radiating cooling of small conducting particles in conditions of the "cutoff" connected with reduction of the sizes of a particle less of wave length of a thermal electromagnetic radiation was considered and essential difference from the formula of the Stefan - Boltzmann [1, 2] is shown. In the present research influence of electric potential of a particle on energy exchange in conditions of homogeneous plasma is added. Effects of energy loss in plasma discharge at presence of metal inclusions were investigated in [3]. The effect considered in the present work can be of interest for research of behavior of a pellet, thrown in hot plasma in the diagnostic purposes [4]. It is shown, that the initial temperature of a particle can define character of consequent temperature dynamics of a particle, and under certain conditions results in to occurrence of two different quasi-stationary thermal conditions which it is possible to consider as existence of *temperature bistability*. Physically such effect is connected with braking an electronic flow at negative charge of a particle and an opportunity of positive charge at elevated temperatures owing to thermal electron emission [5, 6]. It is known, that a local warming up of a surface and as consequence occurrence of strong thermal electron currents can result in such effect as an unipolar arch [5]. In case of the isolated particles thermal electron emission leads only to change of a charge of a particle.

## Model of process of heating

Change of potential of a particle under action of electrons and ions of homogeneous hydrogen plasma on its surface and the current of thermal electron emission considered under the formula of Richardson [5], in approximation of the spherical form is described by the equation (1)

$$C\frac{d\varphi}{dt} = \left\{-en_eV_e\begin{bmatrix}e^{\frac{e\varphi}{kT_e}}, \varphi < 0 \\ 1, \varphi \geq 0\end{bmatrix} + en_iV_i\begin{bmatrix}1, \varphi < 0 \\ e^{-\frac{e\varphi}{kT_i}}, \varphi \geq 0\end{bmatrix} + AT^2e^{-\frac{W}{kT}}\begin{bmatrix}1, \varphi < 0 \\ e^{-\frac{e\varphi}{kT_i}}, \varphi \geq 0\end{bmatrix}\right\} \cdot 4\pi R^2 \qquad (1)$$

where C - electric capacity of a particle, T - temperature of a dust particle, R - its radius, $T_e, n_e, V_e$ - temperature, density and speed of electrons in plasma (similar sizes with an index i correspond to positively charged ions), W – work function of electrons from a surface, A - Richardson's constant [5]. Multipliers in square brackets in both equations define braking electrons and positively charged ions of plasma at approach to a surface of a particle in view of a sign on a charge of a particle. It is supposed, that electrons and ions can only be braked on approach, but not «extracted» from plasma as sign on potential of a surface gets changed – it does not work because of Debye screening of electric fields of a particle in plasma [7].

Time changing of thermal energy of a target in an environment of homogeneous plasma is defined by a flow of high energy electrons and ions from plasma to the surface of a particle whereas energy losses are caused by the thermal radiation, the evaporation of substance, and also by thermal electron emission.

Neglecting time of alignment of temperature in the volume of a ball with radius R~100 μm (here κ- a thermal diffusivity) $\tau_T \sim R^2/4\kappa^2 \sim 10^{-3} s$ these processes can be described by the equation

$$\frac{4}{3}\pi R^3 \rho C_V \frac{dT}{dt} = \left\{ \frac{3}{2}k(T_e - T)n_e V_e \begin{bmatrix} e^{\frac{e\varphi}{kT_e}}, \varphi < 0 \\ 1, \varphi \geq 0 \end{bmatrix} + \frac{3}{2}k(T_i - T)n_i V_i \begin{bmatrix} 1, \varphi < 0 \\ e^{-\frac{e\varphi}{kT_i}}, \varphi \geq 0 \end{bmatrix} - \sigma T^4 - \right.$$

(2)

$$\left. -(H + \frac{3}{2}kT)NC_{snd}e^{-\frac{H}{kT}} - \frac{(W + \frac{3}{2}kT)}{e}AT^2 e^{-\frac{W}{kT}} \begin{bmatrix} 1, \varphi < 0 \\ e^{-\frac{e\varphi}{kT}}, \varphi \geq 0 \end{bmatrix} \right\} 4\pi R^2$$

where c and $C_v$ - density and heat capacity of substance of a target, y - Stefan-Boltzmann constant, H and N - energy of sublimation and density of atoms of a target, $C_{snd}$ - speed of a sound in substance [8]. Besides for simplification of consideration of effects, the radius of a particle was assumed to be constant.

As characteristic time of an establishment of potential of a target
φ_u ~ e( $n_e$ $V_e$ - $n_i$ $V_i$) 4p $R^2$/C ~ 1 ns    at the chosen parameters of plasma $T_e = T_i$ = 0.1-10 eV and $n_e = n_i = 10^{13}$ cm$^{-3}$, on some orders{sequences} is less, than time of an establishment of temperature it is possible to consider an equalization of potential instant and to consider a derivative equal to zero $d\varphi/dt = 0$.
It is possible to find potential in this approach from the second equation at stationary value of temperature, and then substituting in the first one to calculate time dynamics of temperature of a sphere in homogeneous plasma. The method allows to consider also time dependence of parameters of plasma under condition of its local homogeneity.

## Results

Results of modelling of temperature of the carbon ball thrown in hydrogen plasma with density $n_e = n_i = 10^{13}$ cm$^{-3}$ and temperature $T_e = T_i$ = 10 eV, at various initial temperatures of a surface are shown in a Fig. 1. Coefficients for carbon dust pellet were taken from [9, 10]. Change of mass of a ball due to evaporation was not considered. As it is seen from figure, character of change of temperature is defined by an initial warming up of a surface of a ball. At initial temperatures $T_0 \leq$ 3 000 K the temperature of a ball aspires to accept quasistationary value T = 2 600 K. Thus the potential of a particle remains negative and is equal    -36.9 V. If initial temperature $T_0 \geq$ 3 200 K at the same parameters of plasma there is a fast warming up of a particle and temperature aspires to quasi-stationary value T = 4 760 K and with positive potential + 1.8 V. Though change of the sizes of a ball due to evaporation within the frames of the model was not considered, probably nevertheless it is possible to speak about realization of two various modes of heating of a particle in homogeneous plasma (*temperature bistability*) - at least the bottom "cold" quasi-equilibrium state can be reached in experimental conditions, and rise in temperature up to the top "hot" condition will result in to the strengthened evaporation which also can be observed in experiments.

Dependence of a heat flux on a surface of a particle from plasma on surface temperature at various electronic temperatures and density is shown in a Fig. 2. It is seen, that in case of $T_e = T_i = 10$ eV the sign on a heat flux becomes negative between points $2\ 600$ K $\leq T_0 \leq 3\ 100$ K and then one more time at $T_0 \geq 4\ 760$ K. Under negative heat flux there will be a cooling of dust particle that reflects aspiration of temperature to bottom «quasi-equilibrium» value, that dynamics of temperature of particles reflects in a Fig. 1. At lower temperature of plasma $T_e = T_i = 0.1$ and $1.0$ eV only bottom "cold" condition with temperature $1\ 000$ K - $1\ 500$ K can take place. Dependence of a heat flux on surface temperature at various densities of plasma and $T_e = 10$ eV is shown in a Fig. 3.

Dependence of potential of a target on surface temperature in homogeneous plasma at electronic temperatures $T_e = T_i = 0.1$-$10$ eV and density of electrons and ions $n_e = n_i = 10^{13}$ cm$^{-3}$ is shown in a Fig. 4. It is seen that with rise of temperature there a sign of potential of surface of the ball changes, caused by thermal electron emission and the temperature at which the potential changes a sign, corresponds to position of heap on curve of a heat flux density.

In a Fig. 5 the contribution of various processes to a heat flux on a surface of a particle in homogeneous plasma at electronic temperature $T_e = 10$ eV and is shown for density of electrons and ions $n_e = n_i = 10^{13}$ cm$^{-3}$. On the curve 1 all mechanisms of energy transmission are considered, on the curve 2 - radiating losses are rejected, on the curve 3 - cooling due to sublimation is rejected as well. It is seen, that change of a sign of a heat flux in the region of $3\ 000$ K is connected with radiating losses. And the heat losses connected with sublimation of substance of a target become essential at temperatures above $4\ 000$ K. Therefore the warming up of a target to a "hot" quasi-stationary state will result in evaporation of a target. However the "cold" quasi-stationary state can exist long enough time and it is possible to establish an experimental technique on this difference for experimental observation of the effect. At lower temperatures of plasma $T_e \leq 10$ eV the effect of bistability is absent and the "cold" mode is realized only.

## Conclusions

Experimentally the effect of temperature bistability of a dust particle in plasma can be observed by the intensity of a luminescence of a pellet, thrown in plasma with initial temperature above or below the threshold. At parameters of plasma when the temperature bistability is possible higher temperatures of a particle will result in sharp increase of brightness of a luminescence due to evaporation of substance or on the contrary for lower initial temperatures no effect on brightness when the particle tends to cooling. In case of colder and less dense plasma such threshold effect will not be observed.

The author thanks Yu.V.Martynenko for statement of a problem and fruitful discussions.

The work is supported by the Grant of the Russian Fund of Basic Research 06-08-08139 OFI

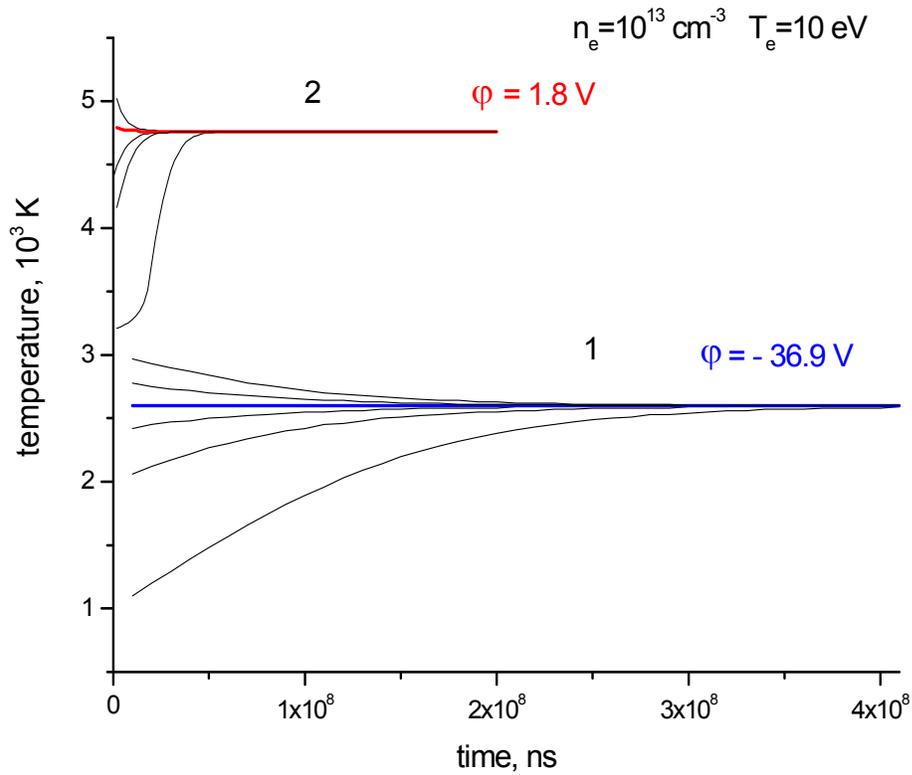

Fig.1.

Time dynamics of temperature of the carbon dust particle thrown in homogeneous plasma with electronic and ionic temperatures $T_e = T_i = 10$ eV and density of electrons and ions $n_e = n_i = 10^{13}$ cm$^{-3}$. Curve group 1 corresponds to initial temperatures of a particle 1 - 3x10$^3$, group 2 - corresponds to initial temperatures 3.2 - 5.4 x10$^3$ K. The radius of a particle relied invariable R=100 мм. Inscriptions near to curves correspond to the positioned potentials of a particle in a "hot" and "cold" quasi-stationary state.

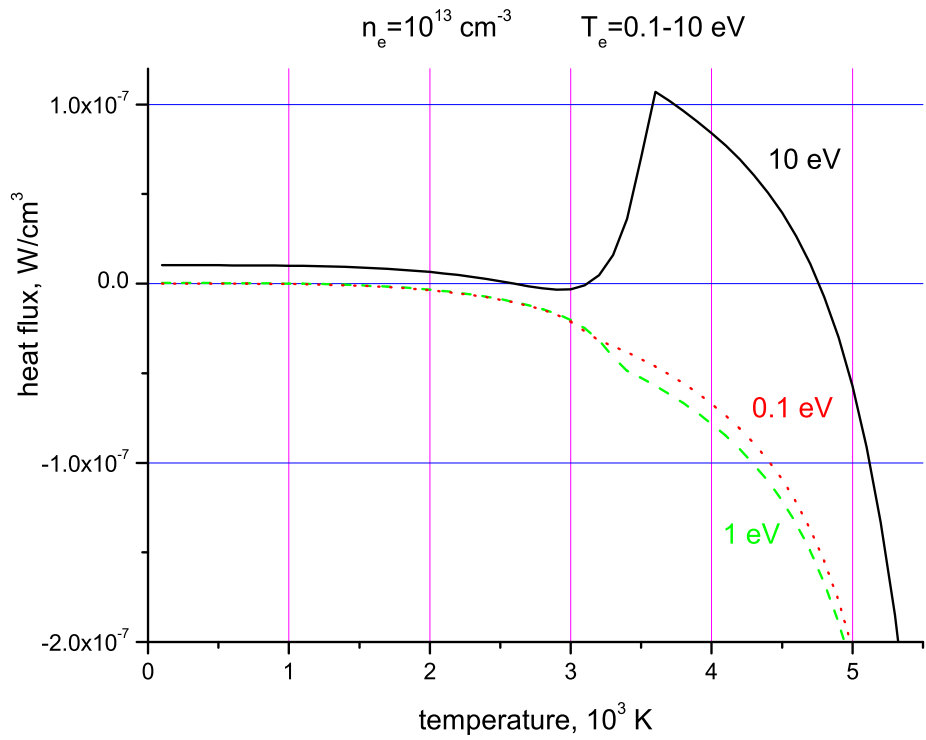

Fig.2.

Dependence of a heat flux on a surface of a particle from plasma from surface temperature at electronic temperatures $T_e = T_i = 0.1\text{-}10$ eV and density of electrons and ions $n_e = n_i = 10^{13}$ cm$^{-3}$.

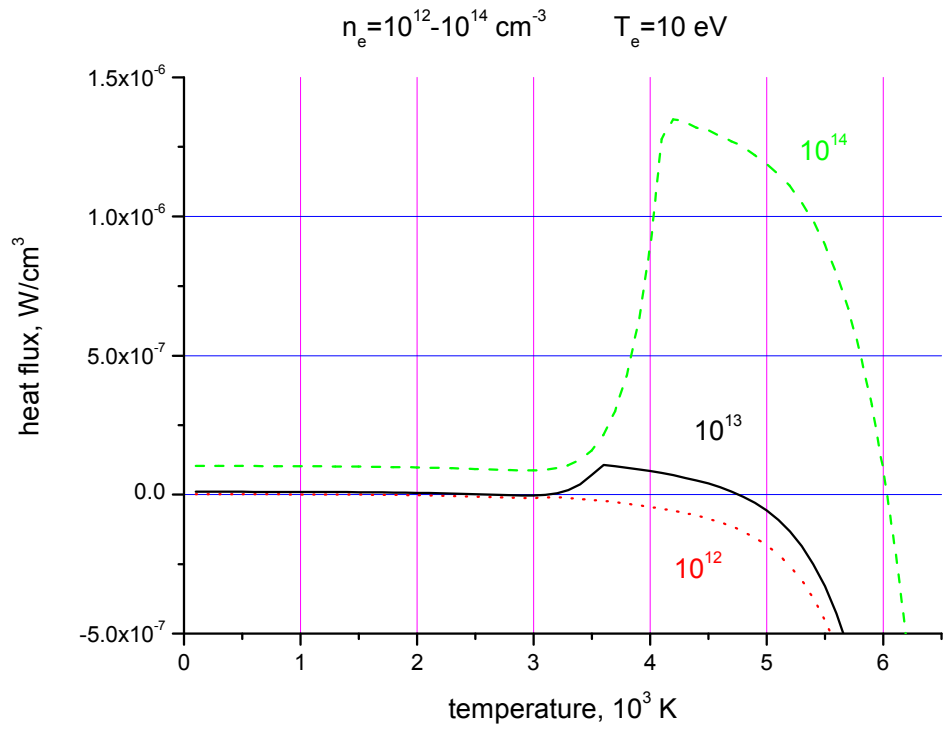

Fig.3.

Dependence of a heat flux on a surface of a particle from plasma from surface temperature at electronic temperature $T_e = T_i = 10$ eV and density of electrons and ions $n_e = n_i = 10^{12} - 10^{14}$ cm$^{-3}$.

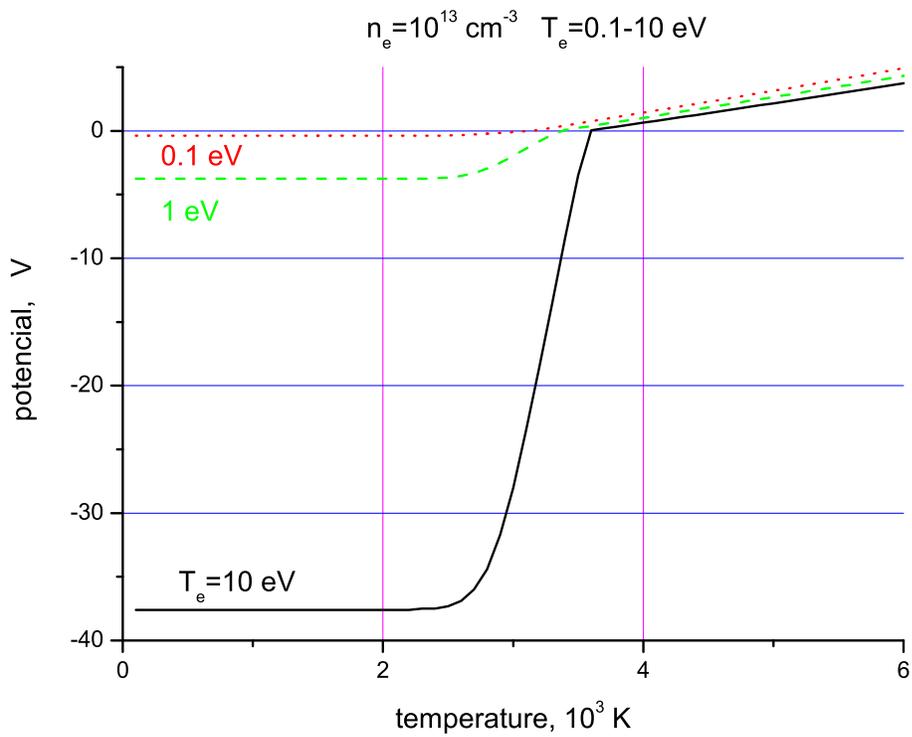

Fig. 4.

Dependence of potential of a target on surface temperature in homogeneous plasma at electronic temperatures $T_e = T_i = 0.1\text{-}10$ eV and density of electrons and ions $n_e = n_i = 10^{13}$ cm$^{-3}$.

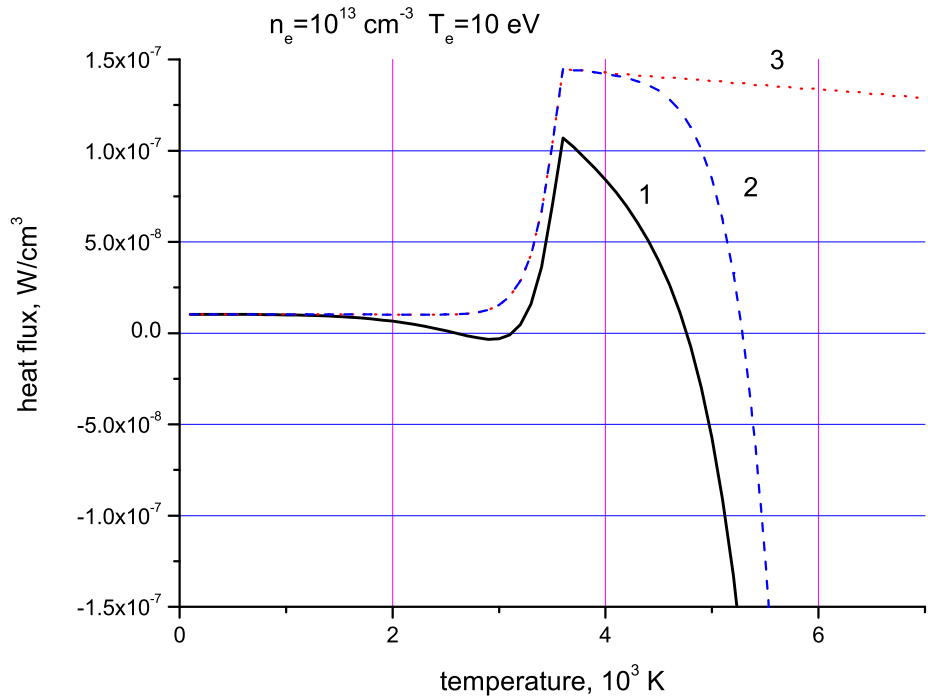

Fig. 5.

The contribution of various processes to a heat flux on a surface of a particle in homogeneous plasma at electronic temperature $T_e$ = 10 eV and density of electrons and ions $n_e = n_i = 10^{13}$ cm$^{-3}$. On the curve 1 the considered mechanisms of energy transmission are considered all, on the curve 2 - radiating losses are rejected, on the curve 3 - cooling due to sublimation and radiation losses is rejected.

.